\documentclass{SCIS2026}

\usepackage{placeins}

\makeatletter
\providecommand{\headerps@out}[1]{}
\makeatother

\graphicspath{{figures/}}


\begin{document}

\ArticleType{RESEARCH PAPER}
\Year{2026}
\Month{}
\Vol{}
\No{}
\DOI{}
\ArtNo{}
\ReceiveDate{}
\ReviseDate{}
\AcceptDate{}
\OnlineDate{}

\title{UniMoE-CP: Unified Conditional Waveform Prediction Across Measured Optical Wireless Links}{UniMoE-CP for Unified Waveform Prediction}

%
%
%
%
%

\author[1,2,3]{Yuning Zhou}{}
\author[1,3]{Ruizhe Jin}{}
\author[1,3]{Yunkai Wang}{}
\author[1,3]{Zhilan Lu}{}
\author[2,4]{Kejia Bian}{}
\author[1,3]{\\Chao Shen}{}
\author[1,3]{Feng Bao}{}
\author[1,3]{Jianyang Shi}{}
\author[1,3]{Ziwei Li}{}
\author[1,3]{Junwen Zhang}{}
\author[1,3]{Nan Chi}{{nanchi@fudan.edu.cn}}

\AuthorMark{Yuning Zhou}
\AuthorCitation{Zhou Y N, Jin R Z, Wang Y K, et al.}

\address[1]{Key Laboratory for Information Science of Electromagnetic Waves (MoE),\\
College of Future Information Technology, Fudan University,\\
Shanghai 200433, China}
\address[2]{Shanghai Innovation Institute, Shanghai 200231, China}
\address[3]{Research Center of Low-Earth-Orbit Satellite Communication and Applications,\\
Shanghai 200433, China}
\address[4]{School of Information Science and Electronic Engineering,\\
Shanghai Jiao Tong University, Shanghai 200240, China}

\abstract{Shared prediction across measured optical wireless links requires a common model to accommodate link-dependent waveform dynamics. This paper studies unified learning for underwater optical, 100-m visible-light communication and 4.3-km free-space optical links. We formulate received-waveform forecasting from historical observations and known future transmitted excitation using one shared parameter set. UniMoE-CP combines a shared pool of temporal experts, window-dependent routing and spectral covariate fusion with an autoregressive decoder. The design permits conditional specialization while reusing expert parameters across the measured links. Pooled-training benchmarks evaluate each link separately and report an aggregated result using normalized mean square error (NMSE) and normalized mean absolute error (NMAE). The reported average NMSE reductions are 29.9\% for 100-m VLC, 29.1\% underwater and 21.4\% for 4.3-km FSO. The aggregated average NMSE decreases from 0.1645 for the best baseline to 0.1178 for UniMoE-CP.}


\maketitle

\section{Introduction}

\begin{figure}[t]
  \centering
  \includegraphics[width=0.84\textwidth]{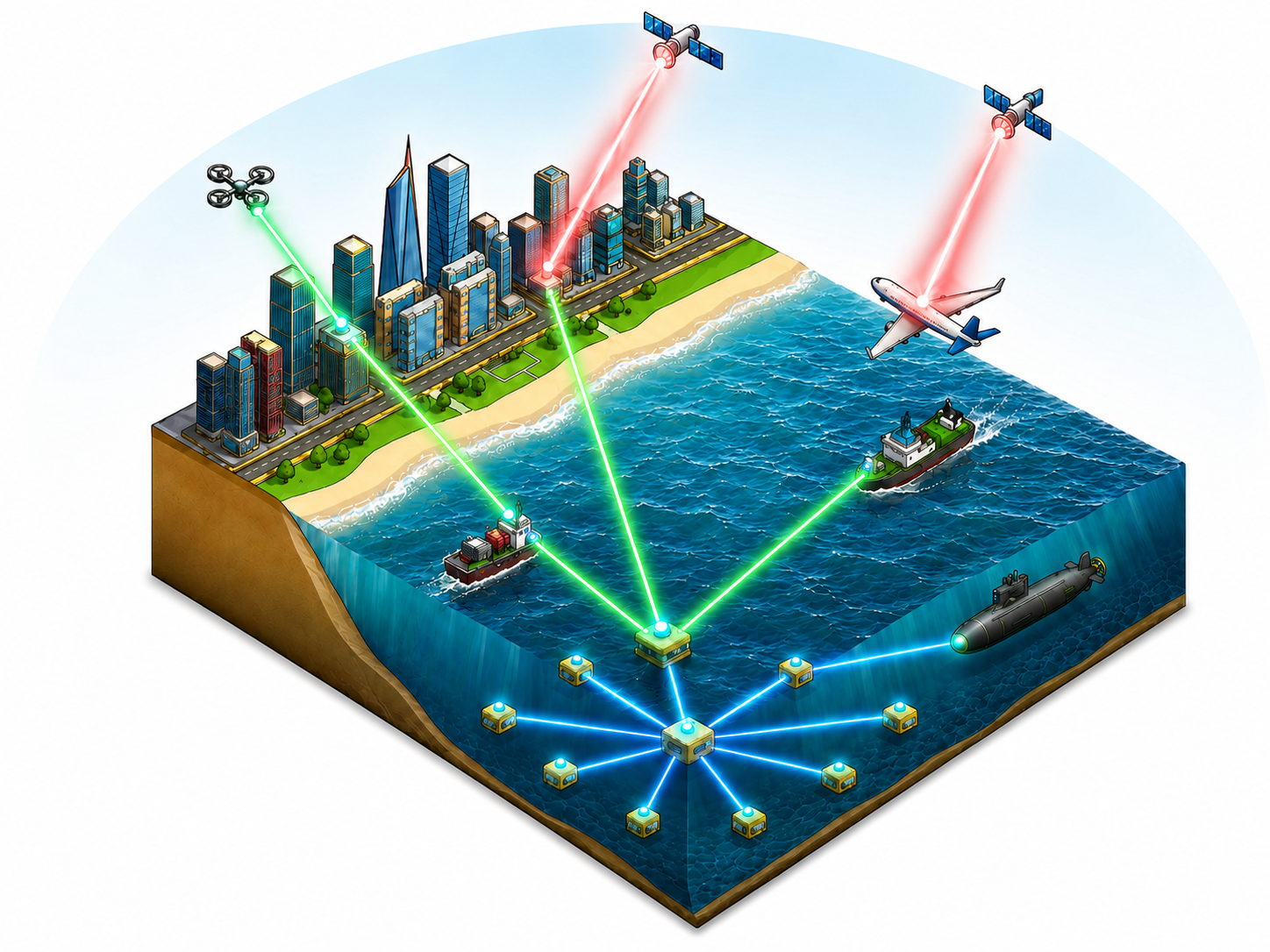}
  \caption{Illustration of an all-scenario optical wireless communication network spanning terrestrial access, air/space-assisted links, maritime surface links and underwater optical sensor links.}
  \label{fig:scene}
\end{figure}

Optical wireless communication (OWC) spans terrestrial access, free-space propagation and underwater links~\cite{chowdhury2020owhn,chi2020vlc6g}. Figure~\ref{fig:scene} illustrates these diverse operating settings. Measurements from such links provide transmitted and received waveforms shaped by both propagation and transceiver responses. Developing a predictor for each measured link creates separate training and model-maintenance workflows. This motivates studying parameter reuse across measured OWC scenarios.

The measured links considered here exhibit different sources of waveform variation. Underwater propagation involves absorption, scattering and medium perturbation~\cite{zeng2017uowcSurvey,lu2025uvllc170g}. Laser-based VLC links involve bandwidth limitation, device nonlinearity and alignment sensitivity~\cite{qin2025tdpcs,liao2026ganpd,zhou2025tbnn}. Outdoor FSO measurements also include atmospheric and beam-coupling fluctuations~\cite{khalighi2014fsoSurvey,liu2025fieldFsoMmw,dong2024multi}. A shared predictor must accommodate these link-dependent statistics together with variation among windows from the same link.

Learning-based channel studies offer several routes toward reusable prediction models. Data-driven OWC simulation and digital-twin-assisted optical modeling provide measurement-oriented approaches~\cite{wei2024owcSimulator,yang2024dt}. LLM4CP adapts a pre-trained language model to historical-to-future CSI prediction using channel-specific input and output processing~\cite{liu2024llm4cp}. WiFo develops unified time-domain and frequency-domain CSI reconstruction across heterogeneous configurations, supported by multi-dataset pre-training and zero-shot evaluation~\cite{liu2025wifo}. Its formulation makes task coverage and model reuse central to the prediction problem.

For measured OWC waveforms, the shared-learning question concerns the relationship between common temporal structure and link-dependent responses. A single parameter set must represent windows with different amplitudes, memory and covariate relations. Conditional specialization provides a way to select temporal components for the current window while retaining a shared pool of parameters. We investigate this design for the three measured links under a common prediction interface.

We define the task as conditional received-waveform forecasting with known transmitted excitation. Each sample contains historical received observations and transmitted covariates, together with the transmitted sequence over the forecast horizon. The predictor outputs future received samples, and its parameters are shared across the underwater, VLC and FSO data. This interface describes measurement-driven prediction under a specified excitation.

To study this setting, we develop UniMoE-CP using a shared temporal expert pool and window-dependent routing. The encoder adapts the temporal and channel clustering modules of DUET~\cite{qiu2025duet}, following sparse expert computation~\cite{shazeer2017moe}. Temporal routing selects decomposition-based features for each waveform window, and spectral gating controls information exchange among waveform variables. An autoregressive LSTM decoder combines the encoded receiver context with the known future transmitted sequence. These components connect parameter sharing to conditional waveform prediction.

\begin{itemize}
\item We formulate a shared-parameter prediction task for heterogeneous measured OWC links. A common history--excitation--response interface specifies the input information and received-waveform target across the three scenarios.
\item We construct a conditionally specialized predictor by combining a shared expert pool, window-dependent temporal routing, spectral covariate fusion and autoregressive waveform decoding. The design reuses encoder parameters across links and allows each input window to select its temporal representation.
\item We evaluate the pooled-training setting on underwater optical, 100-m VLC and 4.3-km FSO measurements using per-link and aggregated NMSE/NMAE. The reported benchmark also examines input history and expert capacity to characterize prediction under the shared parameter set.
\end{itemize}

\section{Unified Conditional Waveform Prediction}

\subsection{Shared prediction interface}

Let $r_t$ denote the received sample at time $t$, and let $\mathbf{u}_t$ denote the transmitted covariates that are available before prediction. Each transmitted covariate vector satisfies $\mathbf{u}_t\in\mathbb{R}^{d_u}$, where $d_u$ is the number of transmitted covariates per prediction step. For the measured-channel window, we use $\boldsymbol{\Omega}_{t}^{\mathrm{in}}$ to collect the historical received waveform and auxiliary covariates. For an input history length $H$ and a forecasting length $L_{\mathrm{out}}$,
\begin{equation}
  \boldsymbol{\Omega}_{t}^{\mathrm{in}}
  =
  [\boldsymbol{\omega}_{t-H+1},\ldots,\boldsymbol{\omega}_{t}]^\top
  \in\mathbb{R}^{H\times N},
  \qquad
  \boldsymbol{\omega}_{\tau}
  =
  [\omega_{\tau,1},\ldots,\omega_{\tau,N}]^\top .
  \label{eq:window_notation}
\end{equation}
The received waveform is one column of this window, indexed by $n_r$, i.e., $\omega_{\tau,n_r}=r_{\tau}$. The remaining columns contain transmitted or auxiliary covariates. The prediction task is expressed as
\begin{equation}
  \widehat{\mathbf{r}}_{t}^{\mathrm{out}}
  =
  \mathcal{F}_{\Theta}
  \left(
  \boldsymbol{\Omega}_{t}^{\mathrm{in}},
  \mathbf{u}_{t+1:t+L_{\mathrm{out}}}
  \right),
  \qquad
  \widehat{\mathbf{r}}_{t}^{\mathrm{out}}
  =
  [\hat r_{t+1},\ldots,\hat r_{t+L_{\mathrm{out}}}]^\top .
  \label{eq:task}
\end{equation}
The parameter set $\Theta$ is shared across all measured links, and training windows use a common input--output layout. The router selects features from a shared expert pool for each input sequence. This permits local temporal choices across windows while reusing the expert parameters. The spectral gate then controls cross-variable fusion before the decoder generates the forecast.

\begin{figure}[t]
  \centering
  \includegraphics[width=0.84\textwidth]{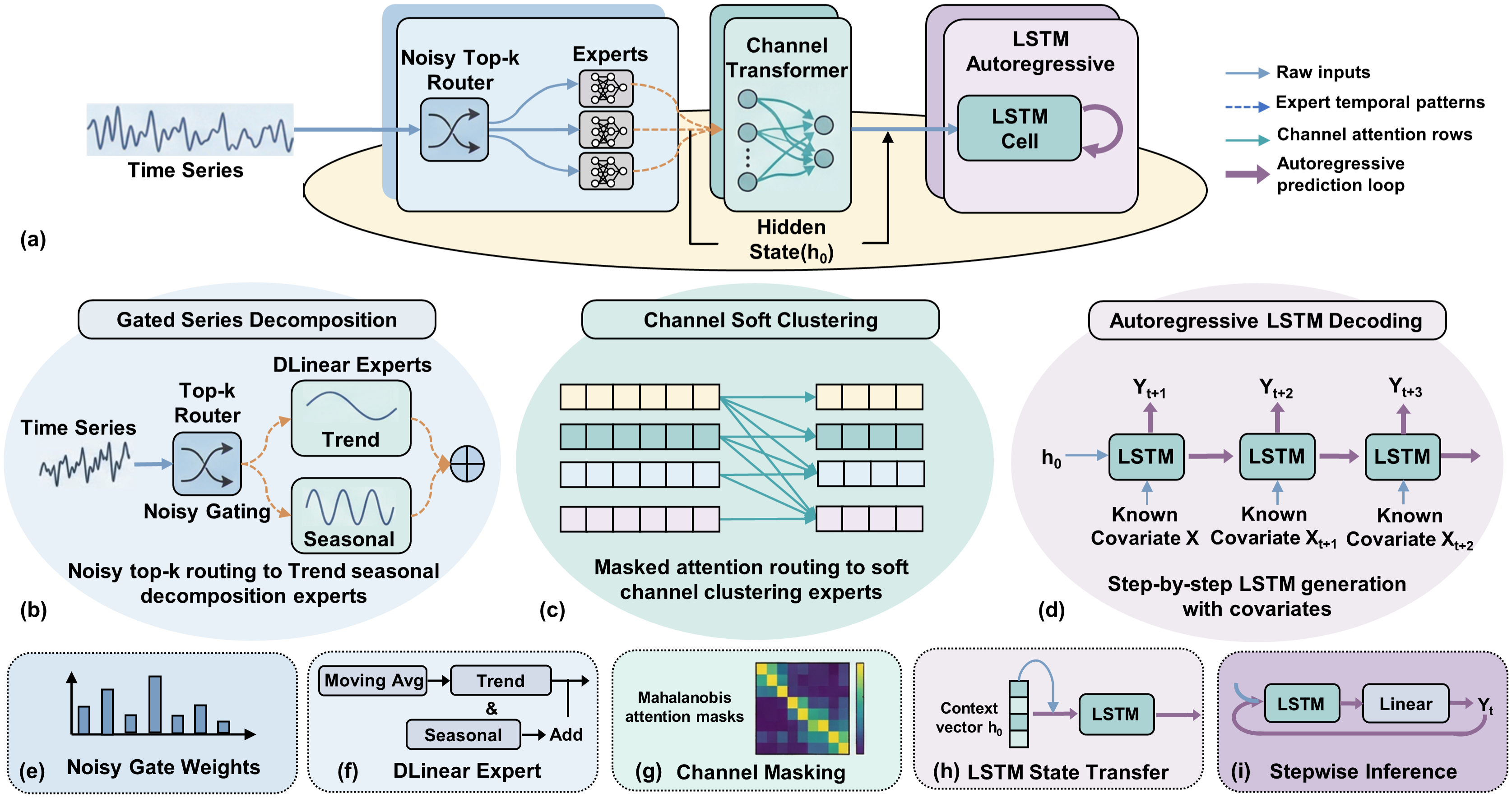}
  \caption{Proposed UniMoE-CP model. (a) Overall encoder-decoder workflow. (b) Gated series decomposition with noisy top-$k$ routing. (c) Channel soft clustering with masked attention routing. (d) Autoregressive LSTM decoding with known future covariates. (e) Noisy gate weights. (f) DLinear expert for slow-fast temporal decomposition. (g) Channel masking in the learned metric space. (h) LSTM state transfer from the encoded context vector. (i) Stepwise autoregressive inference.}
  \label{fig:model}
\end{figure}

Figure~\ref{fig:model} shows the UniMoE-CP architecture. The encoder adapts the temporal and channel clustering modules of DUET~\cite{qiu2025duet}. Temporal-pattern extraction and cross-variable fusion follow the separated structures examined in multivariate forecasting studies~\cite{qiu2024tfb}. An autoregressive LSTM decoder generates received-waveform predictions conditioned on known future transmitted samples. Following the order in Figure~\ref{fig:model}, panel (a) gives the complete encoder-decoder flow, panels (b), (e) and (f) describe temporal routing, panels (c) and (g) describe spectral channel gating, and panels (d), (h) and (i) describe autoregressive waveform generation.

The computation is summarized as
\begin{equation}
  \boldsymbol{\Xi}^{(0)}
  =
  \mathcal{T}_{\psi}
  \left(\boldsymbol{\Omega}_{t}^{\mathrm{in}}\right),
  \qquad
  \boldsymbol{\Xi}^{(J)}
  =
  \mathcal{S}_{\chi}
  \left(\boldsymbol{\Xi}^{(0)},\boldsymbol{\Omega}_{t}^{\mathrm{in}}\right),
  \label{eq:encoder_overview}
\end{equation}
\begin{equation}
  \boldsymbol{\kappa}_{r}
  =
  \boldsymbol{\Xi}^{(J)}_{n_r,:},
  \qquad
  \widehat{\mathbf{r}}_{t}^{\mathrm{out}}
  =
  \mathcal{D}_{\varphi}
  \left(
  \boldsymbol{\kappa}_{r},
  r_t,
  \mathbf{u}_{t+1:t+L_{\mathrm{out}}}
  \right).
  \label{eq:decoder_overview}
\end{equation}
Here $\mathcal{T}_{\psi}$ denotes the temporal-regime routing block, $\mathcal{S}_{\chi}$ denotes the spectral channel-gating block, $J$ is the number of channel-fusion layers, and $\Theta=\{\psi,\chi,\varphi\}$.

\subsection{Temporal-regime routing block}

The temporal block provides conditional specialization within the shared predictor. Each waveform variable is processed by a sparse mixture of experts drawn from the same parameter pool. A window-dependent router assigns the mixture weights according to the observed sequence. The latent regime denotes a temporal representation selected by the model and can be reused by windows from different measured links.

Let $z_n\in\{1,\ldots,E\}$ denote the latent regime of the $n$-th sequence. The routing block approximates the conditional regime distribution by a sparse posterior
\begin{equation}
  p(z_n=e\mid \widetilde{\boldsymbol{\omega}}_n)
  \approx
  \pi_{n,e},
  \qquad
  \sum_{e=1}^{E}\pi_{n,e}=1,
  \qquad
  \|\boldsymbol{\pi}_{n}\|_{0}\leq k .
  \label{eq:sparse_regime_posterior}
\end{equation}
The soft weights determine the mixture within the selected expert set. The cardinality constraint limits each channel sequence to at most $k$ active temporal experts.

For clarity, the batch index is omitted. Let $\boldsymbol{\omega}_{n}\in\mathbb{R}^{H}$ be the $n$-th column of $\boldsymbol{\Omega}_{t}^{\mathrm{in}}$. A reversible normalization is applied as
\begin{equation}
  \widetilde{\boldsymbol{\omega}}_{n}
  =
  \frac{\boldsymbol{\omega}_{n}-a_n\mathbf{1}}
  {b_n+\varepsilon_0},
  \qquad
  a_n=\frac{1}{H}\sum_{\ell=1}^{H}\omega_{n,\ell},
  \qquad
  b_n^2=\frac{1}{H}\sum_{\ell=1}^{H}(\omega_{n,\ell}-a_n)^2 .
  \label{eq:instance_norm}
\end{equation}
where $a_n$ and $b_n$ describe the local offset and scale of the measured sequence. Similar to reversible instance normalization for non-stationary time-series forecasting~\cite{kim2022revin}, this normalization prevents the router from being dominated only by received power or SNR, while the original scale can still be restored after prediction when required. The normalized sequence is then sent to two compact regime encoders:
\begin{equation}
  \boldsymbol{\eta}_{n}
  =
  \mathcal{P}_{a}(\widetilde{\boldsymbol{\omega}}_n),
  \qquad
  \boldsymbol{\chi}_{n}
  =
  \mathcal{P}_{b}(\widetilde{\boldsymbol{\omega}}_n),
  \qquad
  \boldsymbol{\delta}_{n}
  =
  \mathrm{Softplus}(\boldsymbol{\chi}_{n}),
  \label{eq:router_enc}
\end{equation}
where $\boldsymbol{\eta}_{n},\boldsymbol{\chi}_{n},\boldsymbol{\delta}_{n}\in\mathbb{R}^{E}$ and $E$ is the number of temporal experts. Equivalently, the router defines a diagonal Gaussian probe
\begin{equation}
  q_{\psi}(\mathbf{g}_{n}\mid\widetilde{\boldsymbol{\omega}}_n)
  =
  \mathcal{N}
  \left(
  \boldsymbol{\eta}_{n},
  \operatorname{diag}(\boldsymbol{\delta}_{n}^{2})
  \right),
  \label{eq:router_gaussian}
\end{equation}
where the mean represents the deterministic regime evidence and the variance controls how strongly the router explores neighboring experts during training. Using the reparameterization trick, the stochastic regime probe is written as
\begin{equation}
  \boldsymbol{\zeta}_{n}
  =
  \boldsymbol{\eta}_{n}
  +
  \boldsymbol{\xi}_{n}\odot\boldsymbol{\delta}_{n},
  \qquad
  \boldsymbol{\xi}_{n}\sim\mathcal{N}(\mathbf{0},I),
  \label{eq:router_reparam}
\end{equation}
and the expert-selection logits are
\begin{equation}
  \boldsymbol{\ell}_{n}
  =
  R_{\ell}\boldsymbol{\zeta}_{n}
  \in\mathbb{R}^{E}.
  \label{eq:router_score}
\end{equation}
The Gaussian perturbation enables differentiable latent-regime sampling and provides the exploration effect of noisy routing. It is especially useful near regime boundaries: when two experts explain a window with similar confidence, the stochastic probe allows both experts to receive gradients across mini-batches instead of fixing the decision too early.

The relation to noisy routing can be written without duplicating the original formulation. Suppose an alternative noisy selector contains an extra perturbation:
\begin{equation}
  \boldsymbol{\ell}_{n}^{\star}
  =
  R_{1}(\boldsymbol{\eta}_{n}
  +
  \boldsymbol{\xi}_{n}\odot\boldsymbol{\delta}_{n})
  +
  \boldsymbol{\xi}_{n}^{\star}\odot\mathbf{a}_{r},
  \qquad
  \boldsymbol{\xi}_{n}^{\star}\sim\mathcal{N}(\mathbf{0},I).
  \label{eq:noisy_gate_original}
\end{equation}
For its $e$-th component,
\begin{equation}
  \ell_{n,e}^{\star}
  \sim
  \mathcal{N}
  \left(
  (R_{1}\boldsymbol{\eta}_{n})_e,
  \sum_{j=1}^{E}(R_{1,ej}\delta_{n,j})^2+a_{r,e}^{2}
  \right).
  \label{eq:noisy_gate_distribution}
\end{equation}
By contrast, \eqref{eq:router_score} gives
\begin{equation}
  \ell_{n,e}
  \sim
  \mathcal{N}
  \left(
  (R_{\ell}\boldsymbol{\eta}_{n})_e,
  \sum_{j=1}^{E}(R_{\ell,ej}\delta_{n,j})^2
  \right).
  \label{eq:router_distribution}
\end{equation}
Since $R_{\ell}$ is learnable, the variance induced by the compact regime probe can be adjusted through the projection weights, although it is not intended to exactly reproduce every form of independent additive routing noise. The formulation in \eqref{eq:router_reparam} is therefore used as a concise stochastic approximation that retains expert exploration while avoiding an extra noise branch in the routing expression.

Let $\mathcal{I}_{n}=\operatorname{ArgTopK}(\boldsymbol{\ell}_{n},k)$ denote the selected expert-index set. This set is distinct from the Gaussian noise-scale vector $\boldsymbol{\delta}_n$ in \eqref{eq:router_enc}. The sparse routing coefficient is
\begin{equation}
  \pi_{n,e}
  =
  \begin{cases}
  \displaystyle
  \frac{\exp(\ell_{n,e})}
  {\sum_{j\in\mathcal{I}_{n}}\exp(\ell_{n,j})},
  & e\in\mathcal{I}_{n},\\[2mm]
  0, & e\notin\mathcal{I}_{n}.
  \end{cases}
  \label{eq:topk_gate}
\end{equation}
Thus $\sum_e\pi_{n,e}=1$ and only a few experts are activated for each channel sequence.

Each expert uses a slow-fast decomposition. This is motivated by the fact that measured optical wireless waveforms contain both slowly varying channel states, such as received-power drift and memory-like distortion, and faster residual components, such as symbol-scale fluctuations and nonlinear distortion. A moving-average operator separates these two parts:
\begin{equation}
  \boldsymbol{\omega}_{n}^{\mathrm{slow}}
  =
  \mathcal{A}(\widetilde{\boldsymbol{\omega}}_{n}),
  \qquad
  \boldsymbol{\omega}_{n}^{\mathrm{fast}}
  =
  \widetilde{\boldsymbol{\omega}}_{n}
  -
  \boldsymbol{\omega}_{n}^{\mathrm{slow}},
  \label{eq:series_decomp}
\end{equation}
where $\mathcal{A}(\cdot)$ is a moving-average operator. The $e$-th expert then applies two expert-specific projections to the slow and fast components:
\begin{equation}
  \boldsymbol{\psi}_{n}^{(e)}
  =
  \boldsymbol{\omega}_{n}^{\mathrm{slow}}R_{\mathrm{slow}}^{(e)}
  +
  \boldsymbol{\omega}_{n}^{\mathrm{fast}}R_{\mathrm{fast}}^{(e)}
  +
  \mathbf{a}^{(e)},
  \qquad
  \boldsymbol{\psi}_{n}^{(e)}\in\mathbb{R}^{d}.
  \label{eq:expert_output}
\end{equation}
where $R_{\mathrm{slow}}^{(e)},R_{\mathrm{fast}}^{(e)}\in\mathbb{R}^{H\times d}$. The routed temporal code is the posterior-weighted mixture of selected expert outputs:
\begin{equation}
  \boldsymbol{\psi}_{n}
  =
  \sum_{e\in\mathcal{I}_{n}}
  \pi_{n,e}\boldsymbol{\psi}_{n}^{(e)} .
  \label{eq:expert_aggregation}
\end{equation}
An expert represents a temporal basis within the shared pool. Its activation depends on the input window and can occur for sequences from different measured links. Collecting all variables gives the initial token table
\begin{equation}
  \boldsymbol{\Xi}^{(0)}
  =
  [\boldsymbol{\psi}_{1},\boldsymbol{\psi}_{2},\ldots,\boldsymbol{\psi}_{N}]^\top
  \in\mathbb{R}^{N\times d}.
  \label{eq:tcm_output}
\end{equation}

Sparse routing can otherwise collapse to a few dominant experts. To keep experts active, the load-balancing penalty is written with importance $\iota_e$ and activation count $\lambda_e$:
\begin{equation}
  \iota_e=\sum_{b,n}\pi_{b,n,e},
  \qquad
  \lambda_e=\sum_{b,n}\mathbb{I}(e\in\mathcal{I}_{b,n}),
\end{equation}
\begin{equation}
  \mathcal{L}_{\mathrm{bal}}
  =
  \mathrm{cv}^{2}(\boldsymbol{\iota})
  +
  \mathrm{cv}^{2}(\boldsymbol{\lambda}),
  \qquad
  \mathrm{cv}^{2}(\mathbf{a})
  =
  \frac{\mathrm{Var}(\mathbf{a})}{\mathrm{Mean}(\mathbf{a})^2+\varepsilon_0}.
  \label{eq:balance_loss}
\end{equation}
The two terms quantify imbalance in routing probability mass and expert-selection counts, respectively. The activation count $\lambda_e$ is a discrete usage statistic. Gradient-based optimization of this load term requires a differentiable estimator of the selection count.

\subsection{Spectral channel-gating block}

After temporal routing, $\boldsymbol{\Xi}^{(0)}$ contains one token for each variable. Full channel attention is not always desirable because unrelated covariates may introduce noise, while a purely channel-independent model cannot exploit useful transmitter-receiver coupling. UniMoE-CP therefore learns a spectral gate that controls which tokens can exchange information. The goal is to construct a sample-adaptive relation graph
\begin{equation}
  \mathcal{G}_{t}
  =
  \left(
  \mathcal{V},
  \Gamma
  \right),
  \qquad
  \mathcal{V}=\{1,\ldots,N\},
  \qquad
  \Gamma\in[0,1]^{N\times N},
  \label{eq:channel_graph}
\end{equation}
where $\Gamma_{ij}$ measures whether variable $j$ should contribute to the representation of variable $i$ in the current window.

The frequency-domain gate is used as a robust engineering descriptor rather than as a strict signal-reconstruction constraint. Measured optical wireless traces often contain small timing offsets, local phase shifts, slowly varying power drift and bandwidth-limited distortion. Directly comparing two variables sample by sample can therefore overemphasize local misalignment, while comparing their dominant spectral content gives a more stable summary of oscillatory structure and channel-memory behavior. For the $i$-th variable, we use the magnitude of the real fast Fourier transform (rFFT) as the spectral descriptor:
\begin{equation}
  \boldsymbol{\varphi}_{i}
  =
  \left|\mathrm{rFFT}(\widetilde{\boldsymbol{\omega}}_{i})\right|
  \in\mathbb{R}^{L_f},
  \qquad
  L_f=\left\lfloor\frac{H}{2}\right\rfloor+1 .
  \label{eq:rfft_feature}
\end{equation}
The spectral distance is measured by a learnable positive semi-definite metric:
\begin{equation}
  \Delta_{ij}
  =
  (\boldsymbol{\varphi}_{i}-\boldsymbol{\varphi}_{j})^\top
  \Lambda
  (\boldsymbol{\varphi}_{i}-\boldsymbol{\varphi}_{j}),
  \qquad
  \Lambda=B_{\Lambda}^{\top}B_{\Lambda}\succeq 0 .
  \label{eq:mahalanobis}
\end{equation}
Because $\Lambda$ is positive semi-definite,
\begin{equation}
  \Delta_{ij}
  =
  \left\|
  B_{\Lambda}
  (\boldsymbol{\varphi}_{i}-\boldsymbol{\varphi}_{j})
  \right\|_{2}^{2}
  \geq 0 .
  \label{eq:metric_norm}
\end{equation}
Thus the metric learns which frequency components are most informative for downstream channel prediction instead of assuming that all spectral bins contribute equally. This design deliberately uses magnitude-only spectra: phase information is not used to reconstruct the original waveform, but the descriptor weakens sensitivity to small phase shifts and local timing offsets while preserving dominant frequency-content differences needed for channel gating. This is important for optical wireless channels, where low-frequency drift, modulation bandwidth limitation and high-frequency noise may have very different predictive value.

The distance matrix is converted into a probabilistic relationship matrix. With a small $\varepsilon_{\Delta}>0$,
\begin{equation}
  R_{ij}
  =
  \begin{cases}
  \displaystyle\frac{1}{\Delta_{ij}+\varepsilon_{\Delta}}, & i\neq j,\\[2mm]
  0, & i=j,
  \end{cases}
  \label{eq:relation_matrix}
\end{equation}
and
\begin{equation}
  \Pi_{ij}
  =
  \begin{cases}
  \displaystyle
  \gamma\frac{R_{ij}}{\max_{j'\neq i}R_{ij'}+\varepsilon_{\Delta}},
  & i\neq j,\\[2mm]
  1, & i=j,
  \end{cases}
  \qquad
  \gamma\in(0,1).
  \label{eq:prob_matrix}
\end{equation}
Here $\Pi_{ij}$ can be interpreted as the probability that variable $j$ provides useful information to variable $i$. The row-wise maximum normalization expresses each relation relative to the strongest neighbor of the receiving variable. The diagonal of $\Pi$ is set to one before stochastic gate sampling. The discount factor $\gamma$ avoids assigning absolute confidence to cross-channel connections and leaves room for sparsification.

To obtain a sparse gate while preserving gradient propagation, a differentiable Bernoulli relaxation is used. Let $q_{ij}\sim\mathrm{Uniform}(0,1)$ and $\tau_{\Gamma}$ be the relaxation temperature. A soft binary gate can be written as
\begin{equation}
  \Gamma_{ij}
  =
  \sigma
  \left(
  \frac{
  \log \Pi_{ij}-\log(1-\Pi_{ij}+\varepsilon_{\Delta})
  +
  \log q_{ij}-\log(1-q_{ij})
  }{\tau_{\Gamma}}
  \right),
  \label{eq:gumbel_mask}
\end{equation}
where $\sigma(\cdot)$ is the sigmoid function. For fixed $\varepsilon_{\Delta}>0$, the zero-temperature limit has success probability $\Pi_{ij}/(1+\varepsilon_{\Delta})$. For $\Pi_{ii}=1$, this limiting probability is $1/(1+\varepsilon_{\Delta})$. The sampled gate therefore weights information exchange according to the learned spectral relations.

The gate is then used in the fusion attention block. For the $h$-th attention head, the standard query, key and value projections are
\begin{equation}
  Q_h=\boldsymbol{\Xi}^{(j-1)}W_h^Q,\qquad
  K_h=\boldsymbol{\Xi}^{(j-1)}W_h^K,\qquad
  V_h=\boldsymbol{\Xi}^{(j-1)}W_h^V .
\end{equation}
The masked attention logits are
\begin{equation}
  S_{h,ij}
  =
  \frac{\mathbf{q}_{h,i}^{\top}\mathbf{k}_{h,j}}{\sqrt{d_h}}
  +
  \log(\Gamma_{ij}+\varepsilon_{\Gamma}),
  \label{eq:masked_logits}
\end{equation}
where $\mathbf{q}_{h,i}$, $\mathbf{k}_{h,j}$ and $\mathbf{v}_{h,j}$ denote the corresponding rows of $Q_h$, $K_h$ and $V_h$. The additive $\log(\Gamma_{ij}+\varepsilon_{\Gamma})$ term converts the spectral gate into a multiplicative prior in the attention distribution:
\begin{equation}
  \alpha_{h,ij}
  =
  \frac{
  (\Gamma_{ij}+\varepsilon_{\Gamma})
  \exp(\mathbf{q}_{h,i}^{\top}\mathbf{k}_{h,j}/\sqrt{d_h})
  }
  {
  \sum_{j'=1}^{N}
  (\Gamma_{ij'}+\varepsilon_{\Gamma})
  \exp(\mathbf{q}_{h,i}^{\top}\mathbf{k}_{h,j'}/\sqrt{d_h})
  },
  \qquad
  \mathbf{o}_{h,i}
  =
  \sum_{j=1}^{N}\alpha_{h,ij}\mathbf{v}_{h,j}.
  \label{eq:masked_attention}
\end{equation}
Therefore, $\Gamma_{ij}=0$ approximately removes the connection from variable $j$ to variable $i$, while $0<\Gamma_{ij}<1$ weakens but does not completely discard the information path. This soft sparsification is better matched to measured channels than a fixed hard grouping, because cross-channel usefulness can vary with the local signal condition.
After concatenating all heads,
\begin{equation}
  \mathrm{MHA}_{\Gamma}(\boldsymbol{\Xi}^{(j-1)})
  =
  \mathrm{Concat}(\mathbf{O}_{1},\ldots,\mathbf{O}_{H_a})W^O .
\end{equation}
Each fusion layer follows the standard residual attention block:
\begin{equation}
  \widetilde{\boldsymbol{\Xi}}^{(j)}
  =
  \mathrm{LN}
  \left(
  \boldsymbol{\Xi}^{(j-1)}
  +
  \mathrm{MHA}_{\Gamma}(\boldsymbol{\Xi}^{(j-1)})
  \right),
\end{equation}
\begin{equation}
  \boldsymbol{\Xi}^{(j)}
  =
  \mathrm{LN}
  \left(
  \widetilde{\boldsymbol{\Xi}}^{(j)}
  +
  \mathrm{FFN}(\widetilde{\boldsymbol{\Xi}}^{(j)})
  \right).
\end{equation}
The receiver token after $J$ fusion layers,
\begin{equation}
  \boldsymbol{\kappa}_{r}
  =
  \boldsymbol{\Xi}^{(J)}_{n_r,:},
  \label{eq:receiver_token}
\end{equation}
is used as the compact channel-state representation for decoding. The temporal-regime router and the spectral channel gate therefore play complementary roles: the former extracts expert-specific waveform memory, while the latter controls which covariates are allowed to influence the received-signal token.

\subsection{Autoregressive LSTM decoding}

Figure~\ref{fig:model}(d) shows the autoregressive decoder for the known-covariate forecasting task. We define the received-waveform point prediction recursively as
\begin{equation}
\hat r_{t+\tau}
  =
  \mathcal{G}_{\varphi}
  \left(
  \boldsymbol{\kappa}_{r},
  \tilde r_{t:t+\tau-1},
  \mathbf{u}_{t+1:t+\tau}
  \right),
  \qquad \tau=1,\ldots,L_{\mathrm{out}}.
  \label{eq:autoregressive_factorization}
\end{equation}
Here $\mathcal{G}_{\varphi}$ denotes the LSTM decoding map, and $\boldsymbol{\kappa}_{r}$ is the encoded historical context used by this model. The recursive input $\tilde r$ contains the observed initial sample and the preceding decoder inputs defined below. The receiver token is mapped to the initial decoder state through
\begin{equation}
  \mathbf{h}_0=W_h\boldsymbol{\kappa}_{r}+b_h,
  \qquad
  \mathbf{c}_0=W_c\boldsymbol{\kappa}_{r}+b_c.
\end{equation}
At step $\tau$, the known transmitted covariate is $\mathbf{u}_{t+\tau}\in\mathbb{R}^{d_u}$. Concatenating this vector with the previous received value gives
\begin{equation}
  \mathbf{d}_{\tau}
  =
  [\tilde{r}_{t+\tau-1},\mathbf{u}_{t+\tau}]
  \in\mathbb{R}^{1+d_u}.
\end{equation}
A scalar transmitted covariate corresponds to $d_u=1$ and a two-dimensional decoder input.
The state transition and waveform projection are written as
\begin{equation}
  \mathbf{o}_{\tau},(\mathbf{h}_{\tau},\mathbf{c}_{\tau})
  =
  \mathrm{LSTM}
  \left(
  \mathbf{d}_{\tau},
  (\mathbf{h}_{\tau-1},\mathbf{c}_{\tau-1})
  \right),
\end{equation}
\begin{equation}
  \hat{r}_{t+\tau}
  =
  W_o\mathbf{o}_{\tau}+b_o.
\end{equation}
For the first step, $\tilde{r}_{t}=r_t$ is the last observed received sample. For $\tau>1$, training can combine teacher forcing and autoregressive feedback:
\begin{equation}
  \tilde{r}_{t+\tau-1}
  =
  m_{\tau}r_{t+\tau-1}
  +
  (1-m_{\tau})\hat{r}_{t+\tau-1},
  \qquad
  m_{\tau}\in\{0,1\}.
\end{equation}
Here $m_{\tau}=1$ means that the previous ground-truth sample is used, while $m_{\tau}=0$ means that the model feeds back its own prediction. In practice, the probability of $m_{\tau}=1$ can be decreased during training so that the decoder gradually sees its own prediction errors. During inference, $m_{\tau}=0$ for all steps, and the decoder generates the whole horizon recursively as shown in Figure~\ref{fig:model}(i). This design explicitly conditions the future received waveform on both the encoded channel state and the known transmitted sequence.

\subsection{Training objective and metrics}

The prediction loss is computed only on the received-signal column. Since both temporal routing and spectral gating contain stochastic relaxation during training, the empirical objective can be viewed as a mini-batch Monte Carlo approximation of the expected forecasting risk:
\begin{equation}
  \mathcal{L}_{\mathrm{pred}}
  =
  \frac{1}{B L_{\mathrm{out}}}
  \sum_{b=1}^{B}
  \sum_{\tau=1}^{L_{\mathrm{out}}}
  \ell\left(
  \hat{r}_{b,t+\tau},
  r_{b,t+\tau}
  \right).
\end{equation}
where $\ell(\cdot)$ is a point-wise regression loss. The full MoE training objective is written as
\begin{equation}
  \mathcal{L}
  =
  \mathcal{L}_{\mathrm{pred}}
  +
  \lambda\mathcal{L}_{\mathrm{bal}},
\end{equation}
where $\lambda$ controls the auxiliary expert-balancing term. The balancing loss only regularizes the router; the prediction loss remains defined on the physically meaningful received waveform. Performance is reported using NMSE and normalized mean absolute error (NMAE).

The SNR values used for grouping the measured data are estimated from the received constellation rather than assigned as ideal AWGN parameters. Following the EVM-based convention used in the mid-long-distance VLC experiment of Qin et al.~\cite{qin2025tdpcs}, a CAP-demodulated pilot or reference sequence is first synchronized and equalized. Let $s_t=s_{I,t}+j s_{Q,t}$ be the normalized reference constellation symbol, $\tilde{s}_t$ be the corresponding recovered symbol, and $n_t=\tilde{s}_t-s_t=n_{I,t}+j n_{Q,t}$ be the residual error vector. The estimated SNR is then defined as
\begin{equation}
  \mathrm{EstSNR}_{\mathrm{dB}}
  =
  10\log_{10}\left(\frac{1}{\mathrm{EVM}^2}\right)
  =
  10\log_{10}
  \frac{P_0}
  {\frac{1}{T}\sum_{t=1}^{T}
  \left(|n_{I,t}|^2+|n_{Q,t}|^2\right)},
  \label{eq:est_snr}
\end{equation}
where $P_0=\frac{1}{T}\sum_{t=1}^{T}|s_t|^2$ is the normalized reference-constellation power. Thus EstSNR is a post-DSP link-quality estimate that reflects both noise and residual channel distortion, while NMSE and NMAE remain waveform-forecasting errors computed on the future received samples.

\section{Measured Links and Evaluation Protocol}

\subsection{Experimental channel setups}

\begin{figure}[!b]
  \centering
  \includegraphics[width=0.84\textwidth]{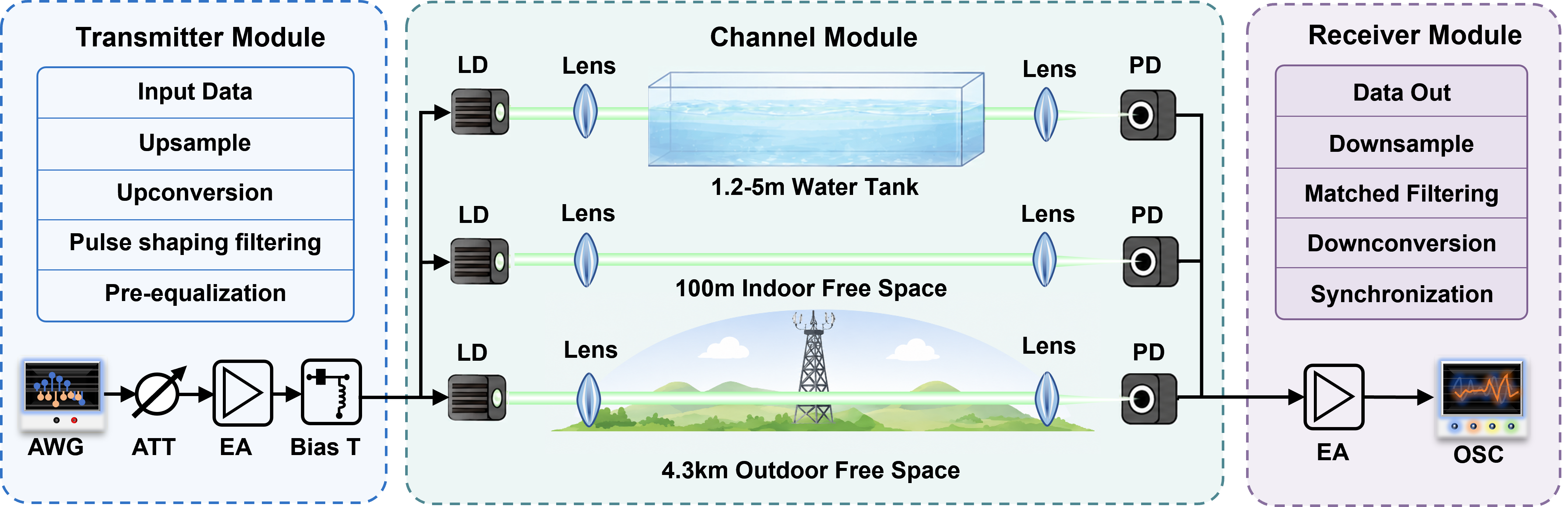}
  \caption{Experimental transmitter-channel-receiver chain for the measured optical wireless links. The left, middle and right parts show the transmitter module, channel module and receiver module, respectively. In the channel module, the top, middle and bottom paths correspond to the underwater optical link, the 100-m VLC free-space link and the 4.3-km outdoor FSO link.}
  \label{fig:system}
\end{figure}

The experiments are conducted on three representative measured optical wireless channels: an underwater optical channel, a 100-m VLC channel and a 4.3-km FSO channel. As shown in Figure~\ref{fig:system}, the three links share a common transmitter-channel-receiver abstraction. The transmitter-side DSP generates the modulated waveform, performs pulse shaping and pre-equalization when needed, and loads the waveform into an arbitrary waveform generator (AWG). The electrical signal is adjusted by amplification, attenuation and DC biasing before driving a laser diode (LD). After optical propagation through the corresponding channel, the received optical signal is converted by a photodetector (PD), sampled by an oscilloscope and processed by receiver-side DSP.

The 100-m VLC channel is established from a mid-long-distance laser-based VLC platform similar to the system reported in~\cite{qin2025tdpcs}. In this setup, the offline transmitter DSP maps the source data to optical wireless waveforms, which are upsampled, pulse-shaped and optionally pre-equalized before AWG output. The electrical waveform is amplified, attenuated and combined with a DC bias to drive the LD. The optical signal propagates through a 100-m indoor free-space path and is then collected by the receiver optics and PD. This channel represents short- to medium-distance access links in which bandwidth limitation, device nonlinearity, alignment sensitivity and received SNR jointly shape the waveform.

The 4.3-km FSO channel is taken from the FSO branch of a 4.3-km urban field demonstration based on a single optical coherent receiver~\cite{liu2025fieldFsoMmw}, and is also related to prior multi-kilometer optical wireless transmission methodology~\cite{dong2024multi}. In the field demonstration, a full-day trial was conducted between Fudan University's Handan Campus and Jiangwan Campus across the urban area of Yangpu District. The FSO optical beam propagated through the outdoor urban path and was collected by the receiver-side optical front end before coherent detection and offline DSP. In this paper, only the FSO received-waveform traces are used, so the 4.3-km data are treated as measured field FSO channel samples rather than as simulated or equivalent-link samples. Compared with the 100-m VLC channel, this scenario emphasizes outdoor turbulence, beam-coupling fluctuation, received-power variation and long-distance free-space channel uncertainty.

The underwater optical channel is constructed using a water-tank visible-light laser communication platform. The transmitter-side waveform is generated by offline DSP and emitted by an LD after electrical amplification and biasing. The optical beam propagates through water before being focused onto the receiver detector. The detected electrical signal is amplified, sampled and processed offline. Compared with air links, the underwater channel is affected by water absorption, scattering, boundary reflection and medium perturbation, which makes it a useful representative of non-air optical propagation environments.

\subsection{Baselines and evaluation protocol}

We compare UniMoE-CP with LSTM~\cite{hochreiter1997lstm}, Transformer~\cite{vaswani2017attention}, LLM4CP, DTnet and CNJCE under the same covariate-conditioned waveform prediction interface. LSTM and Transformer represent recurrent and attention-based sequence predictors; the Transformer baseline is included with awareness that generic Transformer architectures are not always optimal for long-term time-series forecasting~\cite{zeng2023dlinear}. For LLM4CP, which was originally designed for MISO-OFDM CSI prediction with frequency-delay CSI embeddings and a language-model backbone, we replace the CSI-specific embedding with a 1D temporal convolutional embedding so that it can process measured waveform traces. DTnet follows the digital-twin channel modeling idea in~\cite{yang2024dt}, and CNJCE follows the neural OWC channel simulator in~\cite{wei2024owcSimulator}. For DTnet and CNJCE, only the input and output layers are adjusted to match the historical-window input and future received-waveform target.

The benchmark uses a rolling-window, known-covariate forecasting interface. Historical received samples and transmitted covariates form the encoder input, and the future transmitted sequence conditions the decoder. Training windows from the three measured links are pooled to learn one parameter set. Evaluation reports the received-waveform target separately for each link and on the aggregated test set using NMSE and NMAE.

\FloatBarrier
\section{Shared-Parameter Prediction Results}

\subsection{Prediction across measured links}

\begin{figure}[p]
  \centering
  \includegraphics[width=0.78\textwidth]{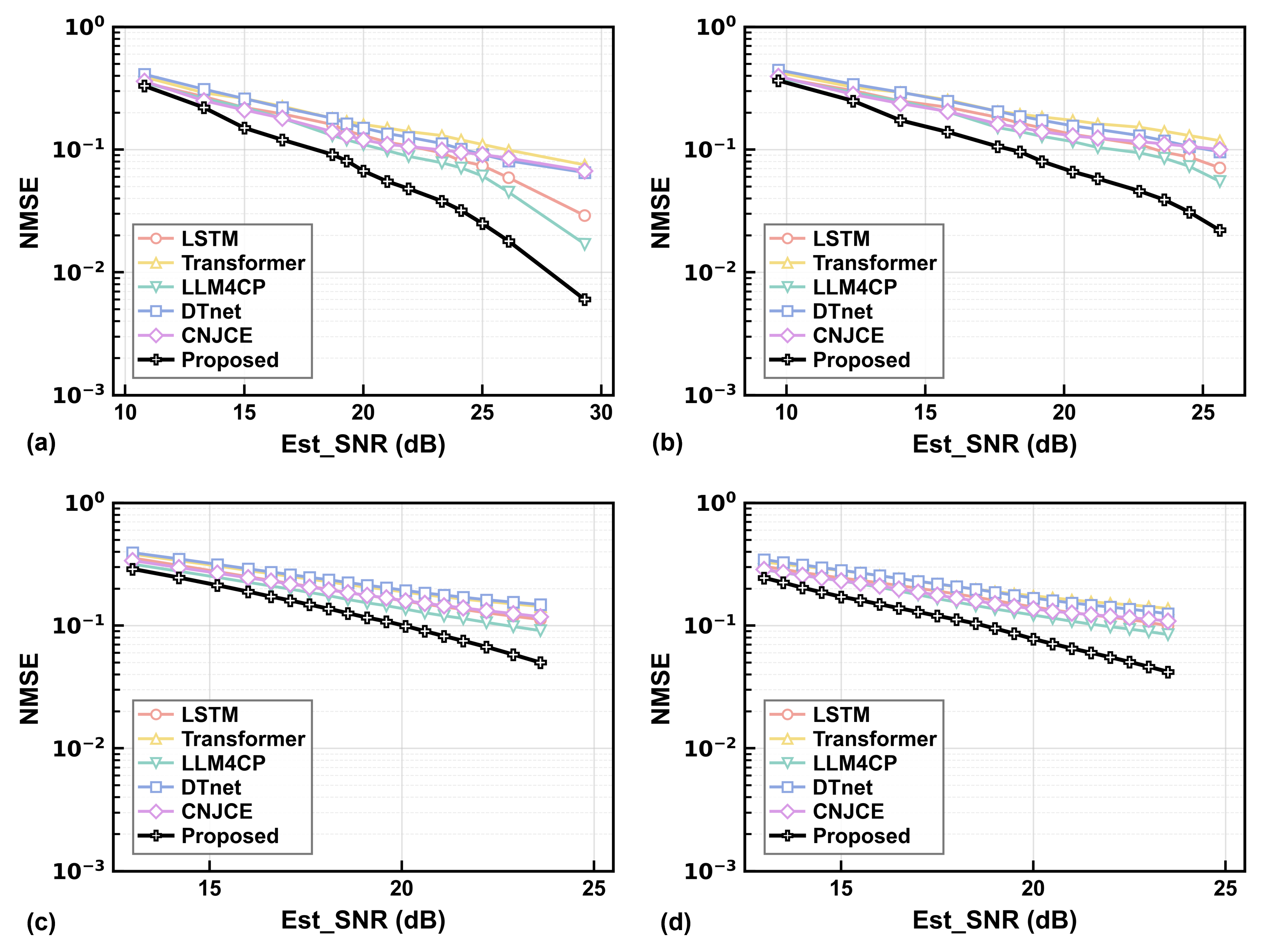}
  \caption{NMSE versus estimated SNR. (a) 100-m VLC free-space link. (b) Underwater optical link. (c) 4.3-km outdoor FSO link. (d) Aggregated evaluation over the three measured links.}
  \label{fig:nmse}
\end{figure}

\begin{figure}[p]
  \centering
  \includegraphics[width=0.78\textwidth]{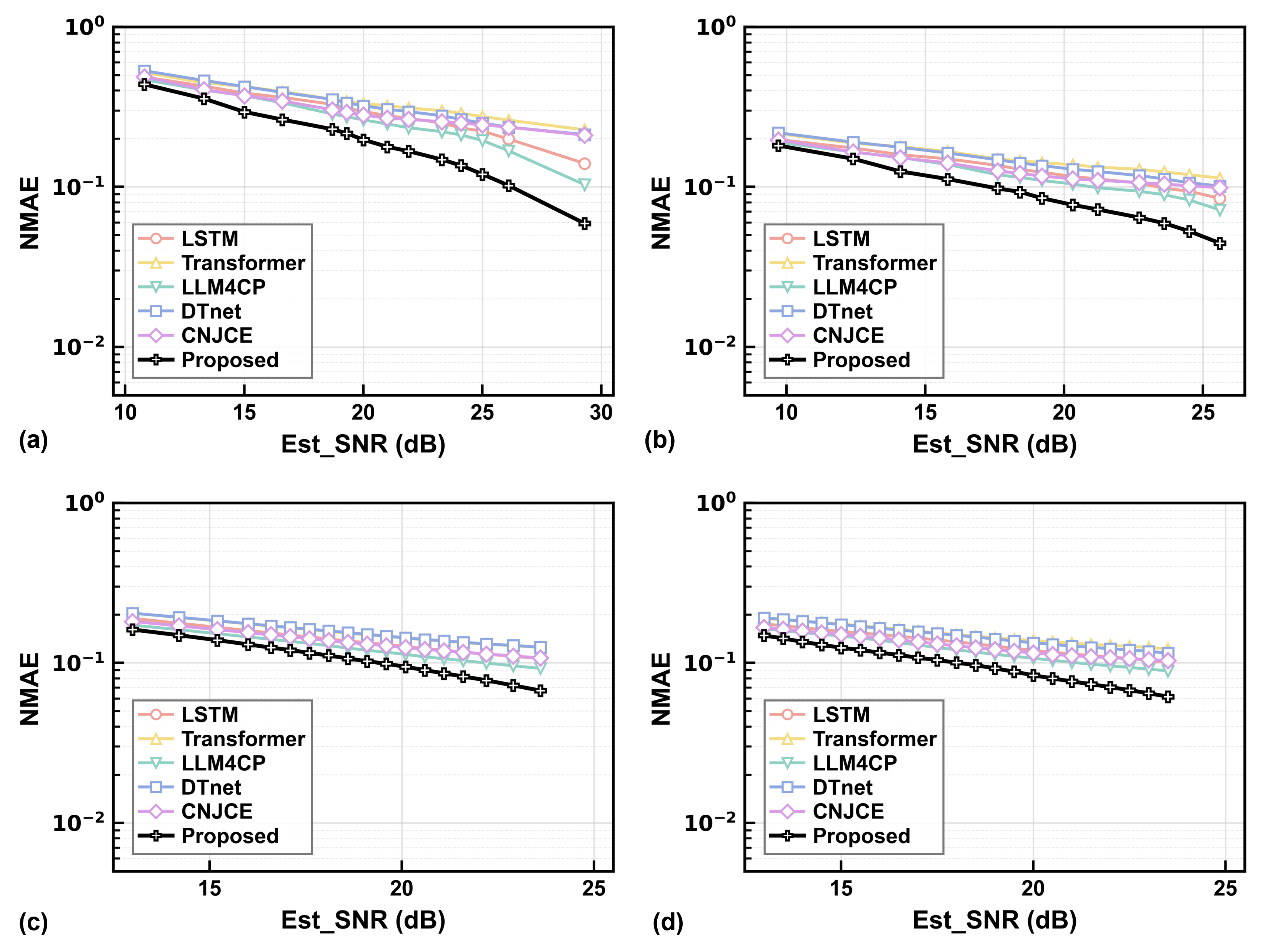}
  \caption{NMAE versus estimated SNR. (a) 100-m VLC free-space link. (b) Underwater optical link. (c) 4.3-km outdoor FSO link. (d) Aggregated evaluation over the three measured links.}
  \label{fig:nmae}
\end{figure}

The primary evaluation examines prediction under a shared parameter set across the three measured links. Figure~\ref{fig:nmse} and Figure~\ref{fig:nmae} report per-link and aggregated errors over the tested estimated-SNR groups. Table~\ref{tab:main_results} summarizes the average benchmark values for each setting.

\begin{table}[H]
  \centering
  \caption{Average performance over all tested estimated-SNR points. The best baseline is reported separately for NMSE and NMAE.}
  \label{tab:main_results}
  \resizebox{0.84\textwidth}{!}{%
  \begin{tabular}{lccccccc}
    \toprule
    Scenario & SNR range (dB) & Proposed NMSE & Best baseline NMSE & NMSE gain & Proposed NMAE & Best baseline NMAE & NMAE gain \\
    \midrule
    100-m VLC & 10.8--29.3 & 0.0915 & 0.1306 (LLM4CP) & 29.9\% & 0.2073 & 0.2697 (LLM4CP) & 23.1\% \\
    Underwater & 9.7--25.6 & 0.1131 & 0.1594 (LLM4CP) & 29.1\% & 0.0933 & 0.1176 (LLM4CP) & 20.6\% \\
    4.3-km FSO & 13.0--23.6 & 0.1347 & 0.1715 (LLM4CP) & 21.4\% & 0.1070 & 0.1247 (LLM4CP) & 14.2\% \\
    Aggregated & 13.0--23.5 & 0.1178 & 0.1645 (LLM4CP) & 28.4\% & 0.0997 & 0.1219 (LLM4CP) & 18.2\% \\
    \bottomrule
  \end{tabular}
  }
\end{table}

For 100-m VLC, underwater optical communication and 4.3-km FSO, the reported average NMSE reductions over the best baseline are 29.9\%, 29.1\% and 21.4\%, respectively. The aggregated NMSE decreases from 0.1645 to 0.1178, corresponding to a reported reduction of 28.4\%. These comparisons characterize the pooled-training benchmark across the measured waveform datasets.

The reported average NMAE reductions are 23.1\% for VLC, 20.6\% underwater and 14.2\% for FSO. The aggregated NMAE reduction is 18.2\%. LLM4CP has the lowest average errors among the waveform-adapted baselines in Table~\ref{tab:main_results}, and UniMoE-CP gives lower reported averages under this interface.

\FloatBarrier

\subsection{Effect of input history}

\begin{figure}[H]
  \centering
  \includegraphics[width=0.84\textwidth]{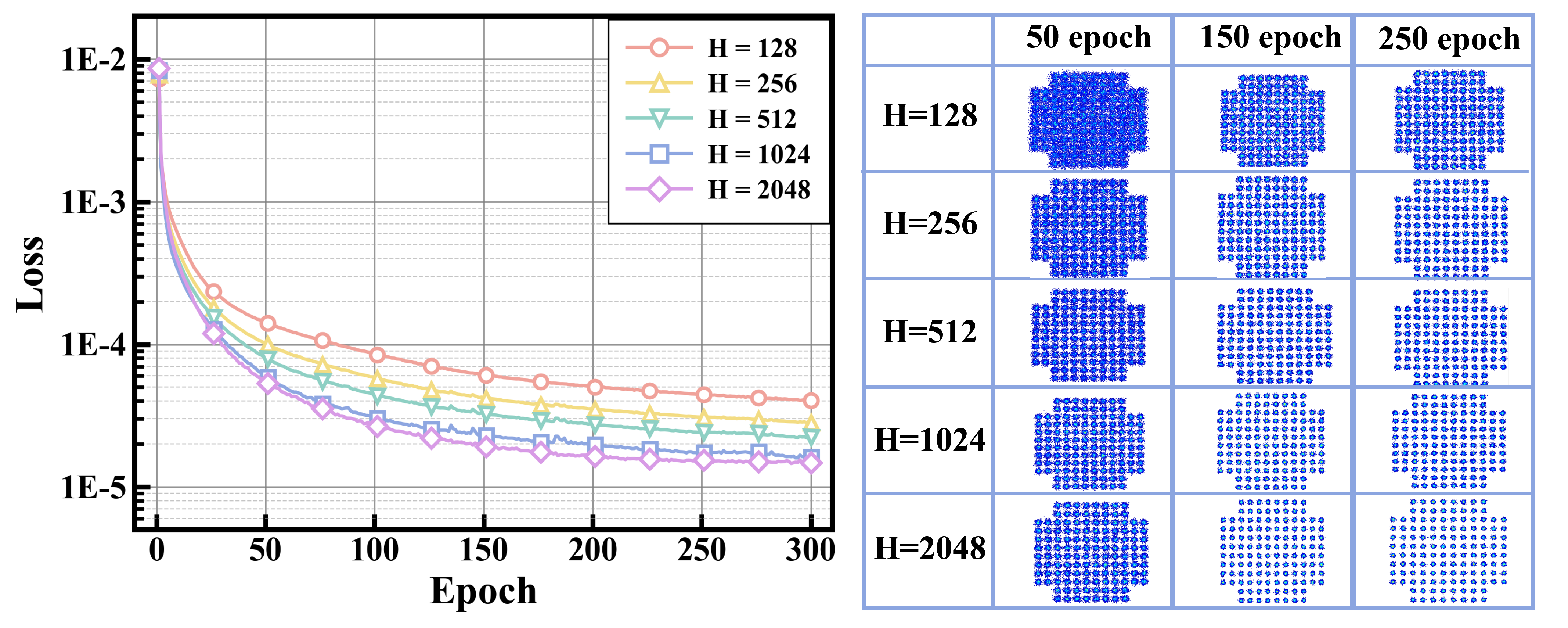}
  \caption{Training loss and constellation snapshots under different input history lengths and training epochs. The left panel shows the training loss versus epoch for different values of $H$. The right panel shows representative recovered constellation snapshots, where different rows correspond to different input history lengths and different columns correspond to selected training epochs.}
  \label{fig:training}
\end{figure}

Figure~\ref{fig:training} examines input history length through the displayed training curves and recovered-constellation snapshots. The final training loss decreases from $4.06\times10^{-5}$ at $H=128$ to $1.47\times10^{-5}$ at $H=2048$. The displayed reduction becomes smaller beyond $H=1024$. This study characterizes the context-length setting used by the predictor.

The constellation panels describe recovered-symbol distributions at the displayed histories and training epochs. Their relationship to waveform-prediction error depends on the receiver DSP and symbol-recovery procedure.

\subsection{Expert capacity under shared parameters}

\begin{figure}[H]
  \centering
  \includegraphics[width=0.84\textwidth]{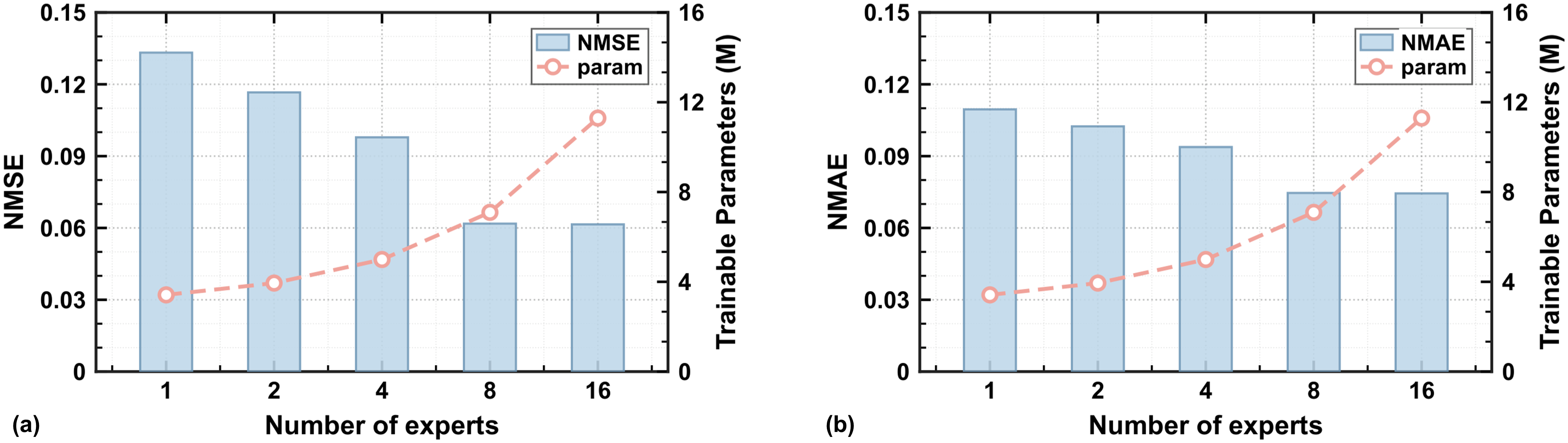}
  \caption{Accuracy and parameter count versus the number of temporal experts. (a) NMSE and trainable parameter count under different expert numbers. (b) NMAE and trainable parameter count under different expert numbers.}
  \label{fig:experts}
\end{figure}

Figure~\ref{fig:experts} examines conditional capacity by varying the number of temporal experts within the shared predictor. Increasing the expert count from one to eight changes the reported NMSE from 0.1332 to 0.0618 and NMAE from 0.1095 to 0.0746. With 16 experts, the error reduction relative to eight experts is small, while the parameter count increases from 7.09 M to 11.30 M.

The eight-expert setting is used for the main benchmark. Routing selects at most $k$ experts per input sequence, while the complete model retains the full expert pool. The plotted study therefore records an accuracy--parameter tradeoff within the shared architecture. Runtime and memory requirements also depend on variable count, forecast horizon and decoder execution.

\subsection{Waveform display pending provenance verification}

\begin{figure}[H]
  \centering
  \includegraphics[width=0.84\textwidth]{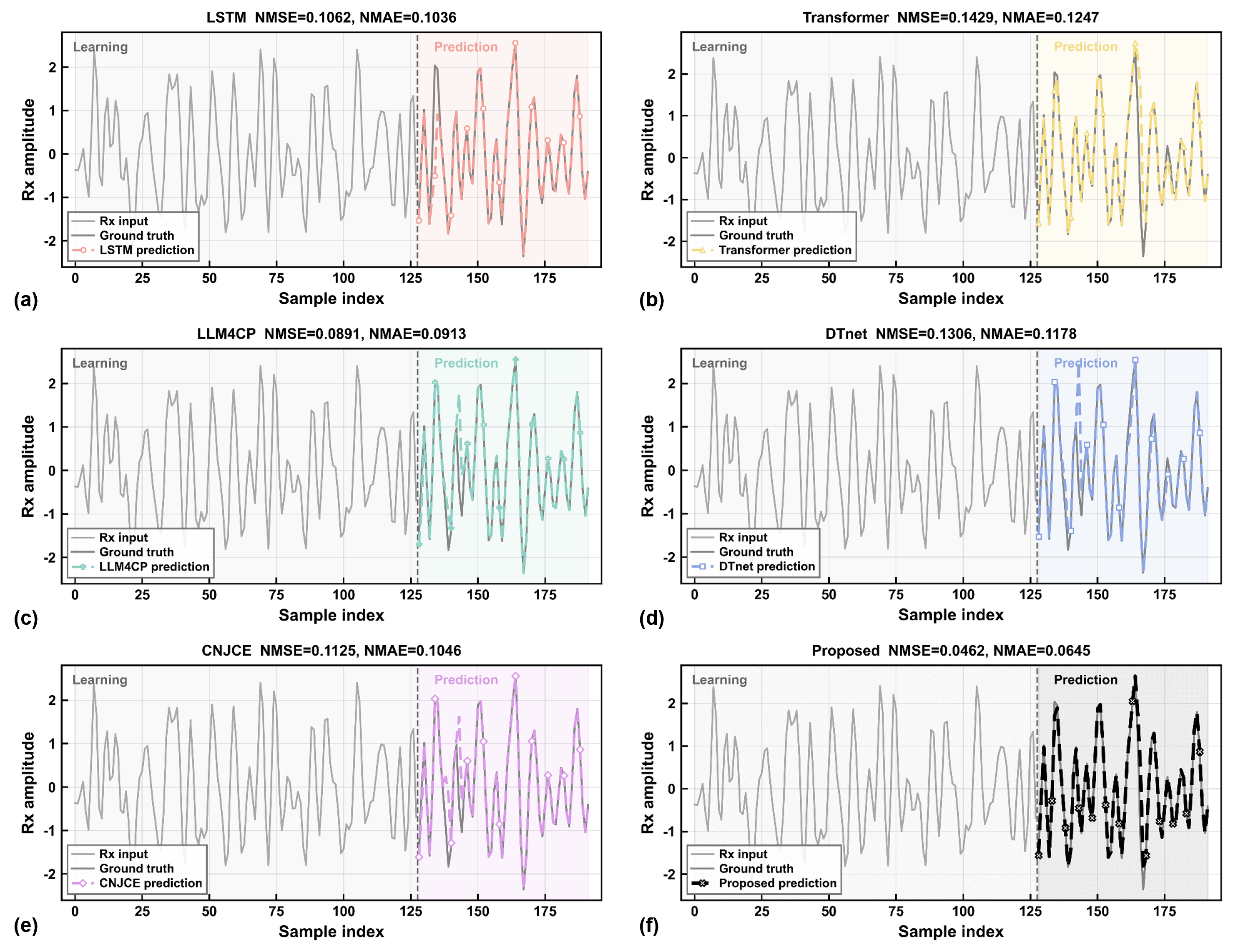}
  \caption{Waveform display at estimated SNR 23 dB. (a) LSTM. (b) Transformer. (c) LLM4CP. (d) DTnet. (e) CNJCE. (f) UniMoE-CP. The shaded region is the forecast horizon. \emph{Draft provenance note. The plotted sequences require verification against saved model outputs.}}
  \label{fig:prediction}
\end{figure}

Figure~\ref{fig:prediction} retains the waveform display associated with the estimated-SNR 23 dB case. It separates the historical observation window from the forecast window and labels the compared methods. The plotted sequences require verification against saved model outputs before supporting pointwise performance conclusions.

\FloatBarrier

\section{Conclusion}

This paper studied unified conditional waveform prediction across heterogeneous measured optical wireless links. UniMoE-CP uses a shared temporal expert pool with window-dependent routing and spectral covariate fusion. Its autoregressive decoder predicts received samples using historical context and known future transmitted excitation. The formulation connects parameter reuse across links to conditional temporal specialization within each window.

The reported pooled-training benchmark covers underwater optical, 100-m VLC and 4.3-km FSO measurements. Per-link and aggregated NMSE/NMAE characterize prediction using one parameter set, and the expert-count study examines conditional capacity. Quantifying the gain from shared training requires controlled comparisons with scenario-specific models. Held-out-link evaluation and adaptation provide further tests of model reuse across new measured conditions.


\Acknowledgements{This work was supported by the Natural Science Foundation of China Project (No. 62525102).}

\bibliographystyle{scis}
\bibliography{references_scis2026}

\end{document}